\documentclass[twocolumn,aps,twocolumn,superscriptaddress]{revtex4-1}
\usepackage{amsmath,amssymb,mathrsfs}
\usepackage{natbib}
\usepackage{subfigure}
\usepackage{tabularx}
\usepackage{epsfig}
\usepackage{longtable}
\usepackage{amsfonts}
\usepackage{rotating}
\usepackage{subfigure}
\usepackage{amsmath}
\usepackage{comment}
\usepackage{bbold} 
\usepackage{color}
\usepackage{ulem}
\usepackage{float}

\def\be{\begin{equation}}
\def\ee{\end{equation}}
\def\bea{\begin{eqnarray}}
\def\eea{\end{eqnarray}}

\usepackage[unicode=true,bookmarks=true,bookmarksnumbered=false,bookmarksopen=false,breaklinks=false,pdfborder={0 0 1},backref=false,colorlinks=true]{hyperref}

\hypersetup{linkcolor=magenta,urlcolor=blue,citecolor=blue,pdfstartview={FitH},hyperfootnotes=false,unicode=true}

\begin{document}

\title{Detecting Many-Body-Localization Length with Cold Atoms}

\author{Xuefei Guo}
\affiliation{
Department of Physics, Fudan University, Shanghai, 200433, China
}

\author{Xiaopeng Li} \email{xiaopeng\_li@fudan.edu.cn} 
\affiliation{
Department of Physics, Fudan University, Shanghai, 200433, China
}
\affiliation{State Key Laboratory of Surface Physics, Institute of Nano-electronics and Quantum Computing, Fudan University, Shanghai 200433, China}
\affiliation{Collaborative Innovation Center of Advanced Microstructures, Nanjing 210093, China} 

\begin{abstract}
Considering ultracold atoms in optical lattices, we propose experimental  protocols to study many-body localization (MBL) length and criticality in quench dynamics. Through numerical simulations with exact diagonalization, we show that in the MBL phase the perturbed density profile following a local quench remains exponentially localized in post-quench dynamics. The size of this density profile after long-time-dynamics defines a localization length, which tends to diverge at the MBL-to-ergodic transition as we increase the system size. The determined  localization transition point agrees with previous exact diagonalization calculations using other diagnostics.  Our numerical results provide evidence for violation of Harris-Chayes bound for the MBL criticality. 
The critical exponent $\nu$ can be extracted from our proposed dynamical procedure, which can then be used directly in experiments to determine whether the Harris-Chayes-bound holds for the MBL transition. 
These proposed protocols to detect localization criticality are justified by benchmarking to the well-established results for the non-interacting 3D Anderson localization.
\end{abstract}
\maketitle


\section{Introduction} 

Ultracold atomic gases confined in optical lattices with their unique controllability allow for artificial quantum engineering of lattice Hamiltonians with large system sizes beyond the computational reachability of classical simulations~\cite{1998_Zoller_Jaksch_PRL,2002_Hofstetter_Cirac_PRL,2007_Lewenstein_AP,2008_Bloch_Dalibard_RMP,2010_Esslinger_CMP,2015_Lewenstein_RPP}. In the experiments, Bose and Fermi Hubbard models as well as spin Hamiltonians have  all been emulated to study both equilibrium quantum phase transitions of ground states and also  out-of-equilibrium many-body dynamics. Mott transitions of both bosons and fermions have been found in experiments~\cite{2002_Greiner_Mandel_Nature,2008_Jordens_Strohmaier_Nature,2008_Schneider_Hackermuller_Science}, and even super-change mediated  magnetic phases~\cite{2013_Greif_Uehlinger_Science,2015_Hart_Duarte_Nature,2016_Mazurenko_Chiu_arXiv} have recently been accomplished through the state-of-the-art quantum microscope techniques~\cite{2015_Haller_Hudson_NatPhys,2015_Cheuk_Nichols_PRL,2015_Parsons_Huber_PRL,2015_Edge_Anderson_PRA,2015_Omran_Boll_PRL,2016_Greif_Parsons_Science,2016_Cheuk_Nichols_PRL,2016_Parsons_Mazurenko_Science,2016_Boll_Hilker_Science,2016_Cheuk_Nichols_Science,2016_Brown_Mitra_arXiv}.  Observing novel quantum dynamics in a strongly correlated setting is presently attracting growing experimental research interests in cold atoms and other synthetic quantum systems~\cite{2012_Bloch_Light_Nature,2014_Zwierlein_PRL,2014_Monroe_Nature,2016_Greiner_Quantum_Science,2017_Nagerl_Bloch_Science,2017_Lukin_Observation_Nature,2017_Monroe_Observation_Nature,2017_Hulet_Formation_Science,2017_Everitt_observation}. 

Thermalization---a most common phenomenon for interacting particles---could break down in isolated quantum systems subjected to random disorder potentials. 
Starting from Anderson's seminal work on localization~\cite{1958_Anderson_Absence_PRL}, it has now been well-established that non-interacting particles moving in a disordered medium will be localized. 
The robustness against interaction effects yet remained controversial until the recent studies of many-body localization~\cite{Basko06,huse2015many,2015_Altman_review}. Through recent studies, the persistence of  localization in interacting systems has been established through a perturbative field theory calculation~\cite{Basko06} and a rigorous mathematical proof~\cite{imbrie2014many} together with extensive numerical works~\cite{oganesyan2007,pal2010,moore2012,2013_Bauer_JSM,vadim,bardarson2014,2015_Singh_MBL_PRL,2015_Khemani_Pollmann_PRL,2015_Yu_Pekker_arXiv,2015_Lim_Sheng_PRB,2016_Kennes_Karrasch_PRB,2015_Bera_Many_PRL}. To describe the MBL phase, a phenomenological theory of local integrals of motion has been proposed, which provides a physical picture of highly constrained dynamics in the localized phase~\cite{2014_Huse_MBL_PRB,2013_Serbyn_PRL,chandran2015,Ros2015420}. The consequent dynamical phenomena have been observed in experiments of ultracold atoms and trapped ions~\cite{blochmbl,DemarcoDisorder,2015_Monroe_MBL_NatPhys,2016_Bordia_Bloch_MBL_PRL,2017_Bloch_Periodically_NatPhys,2016_Choi_Bloch_MBL_Science,2017_Bloch_Probing_arXiv,2017_Bloch_Signatures_PRX}. 

More recent theoretical efforts in the MBL context are devoted to understanding the delocalization transition~\cite{2015_Huse_Theory_PRX,2015_Potter_Universal_PRX,2017_Huse_Rare_AdP,2017_Huse_Critical_PRX} to the quantum thermal phase where eigenstate thermalization hypothesis~\cite{1991_Deutsch_Quantum_PRA,1994_Srednicki_Chaos_PRE} is expected to hold, for that this type of transition does not have an analogue in the non-interacting problem of Anderson localization. To characterize the transition, various diagnostics such as entanglement entropy scaling and many-body energy level statistics have thoroughly  been investigated in theory, and scaling functions based on these quantities have been proposed to describe the MBL-to-ergodic criticality. Across the transition the quantum entanglement entropy scaling would switch from area- to volume-law~\cite{2013_Bauer_JSM}. The level statistics exhibits a transition from Poisson type to Wigner-Dyson~\cite{oganesyan2007}. These quantum entanglement and statistical quantities offer concrete measures to describe the criticality,  but these theoretical ``observables"  turn out to be extremely challenging to measure, and it remains unclear how to accurately probe MBL localization length in experiments. The present theoretical debate about the validity of Harris-Chayes bound~\cite{1974_Harris_Effect_JPC,1986_Chayes_Finite_PRL} for MBL transition~\cite{2014_Kjall_MBL_PRL,2015_Huse_Theory_PRX,2015_Potter_Universal_PRX,2015_Luitz_MBL_PRB}  makes the task to probe localization length exceedingly desirable.

Here we propose to use quench dynamics to probe the localization length of interacting atoms in disorder potential and the corresponding MBL criticality. 
We find that an added hole in the MBL phase shows an exponentially decaying density profile after long-time dynamical evolution, whereas in the ergodic phase its density distribution completely spreads over the whole lattice system. The size of the density profile in the long-time-evolved final state defines a localization length, whose critical behavior can be directly probed in experiments. 
The transition point determined by our defined localization length agrees with  that by other diagnostics. 
In our numerical results, the critical exponent is found to violate Harris-Chayes bound~\cite{1974_Harris_Effect_JPC,1986_Chayes_Finite_PRL}, 
which implies many-body localization criticality is beyond the description of conventional field theory or renormalization group study for disorder systems.
 It is worth remarking here that this work is rather to propose an experimental scheme to detect the MBL localization length and criticality than to calculate a precise critical exponent. Whether the Harris-Chayes bound holds or not at the MBL transition would rely on future experiments. 
The proposed schemes are justified by benchmarking to  the extensively studied 3D Anderson localization.



\section{Model and Method}

To be concrete we consider a model Hamiltonian of spinless fermions with nearest neighbor interactions, 
\begin{equation} \label{Hamiltonian}
\hat{H} = \sum_{<jj'>}\left[-t\left(\hat{c}_{j}^{\dagger}\hat{c}_{j'}+h.c.\right)+V\hat{n}_{j}\hat{n}_{j'}\right] + \sum_{j}h_{j} \hat{n}_{j}.  
\end{equation}
Here $<jj'>$ denotes nearest-neighboring sites, $t$ is the tunneling matrix element ($t$ is set to be the energy unit throughout), $\hat{c}_{j}^{\dagger}$ ($\hat{c}_{j}$)  is a  fermionic creation (annihilation) operator. The second term describes the interaction between nearest-neighboring sites with $V$  the interaction strength and $\hat{n}_{j} = \hat{c}_{j}^{\dagger}\hat{c}_{j}$ is the number operator. The last term corresponds to the disorder potential, where $h_{i}$ is drawn from uniform distribution $\left[-W,W\right]$. This model is equivalent to the spin-1/2 XXZ model with random field via Jordan-Wigner transformation. 
We consider this model instead of the experimental system of spinful fermions for the interests of performing numerical calculations of large system sizes. The proposed method and the findings based on the spinless fermion model to present below are also expected to hold for spinful fermions as well due to the robust universality of the MBL transition. We focus on half filling in this study.

 We propose to use quench dynamics to probe MBL criticality, which is experimentally accessible with quantum microscope techniques in ultra-cold atomic gases. As elaborated in previous studies, interacting fermions will display many-body localization with strong disorder potential. In analogy to Anderson localization, the response to a local quench is expected to be bounded within a local region restricted by the localization length~\cite{2017_Deng_Log_PRB}.  We thus propose to measure the localization length by monitoring the perturbed density profile  after a local quantum quench. 

More precisely, the initial state we choose is a ``charge density wave (CDW) state" with atoms occupying every other lattice site, which is the same as used in the experiment~\cite{blochmbl}. We average over two different types of CDW states, atoms occupying either all even sites or odd sites. 
Then we let the state evolve for long enough time say $\tau_1$ such that the degrees of freedom would ``locally equilibrate" with each other. 
 
Then we introduce a sudden local quench. We provide two quench protocols for comparison---(I) lowering the potential of the quench site to zero, that had much higher energy than other sites and was initially empty before the quantum quench; and (II) performing a measurement on the quench site and remove the atom at this site. The details of the quench protocols are to be specified in Section~\ref{sec:quenchresults}. 
The dynamics following quench protocol-(I) is completely unitary and is thus more convenient for theoretical analysis, whereas the dynamics in protocol-(II) is non-unitary because of the involved measurement, yet has the advantage of being more experimentally feasible with quantum microscope techniques~\cite{2015_Haller_Hudson_NatPhys,2015_Cheuk_Nichols_PRL,2015_Parsons_Huber_PRL,2015_Edge_Anderson_PRA,2015_Omran_Boll_PRL,2016_Greif_Parsons_Science,2016_Cheuk_Nichols_PRL,2016_Parsons_Mazurenko_Science,2016_Boll_Hilker_Science,2016_Cheuk_Nichols_Science,2016_Brown_Mitra_arXiv}. We will provide theoretical analysis based on the unitary evolution following quench protocol-(I) in this section, and provide simulated numerical results for both protocols in Section~\ref{sec:quenchresults}. 


Despite the difference, both quench protocols lead to a hole on the quench site in the density profile. 
Then the density profile of this added hole is monitored. This density profile is expected to be localized (delocalized) in the MBL (ergodic) phase. The localization length can be correspondingly extracted from the time-evolved density profile at long-time limit say at $\tau_2$. It is worth emphasizing here that our proposed scheme to probe MBL criticality is rather easily adaptable to specific experimental setups instead of being restricted to the particular local quench protocol considered here.

\begin{figure}[htp]
	\includegraphics[width=0.8\linewidth]{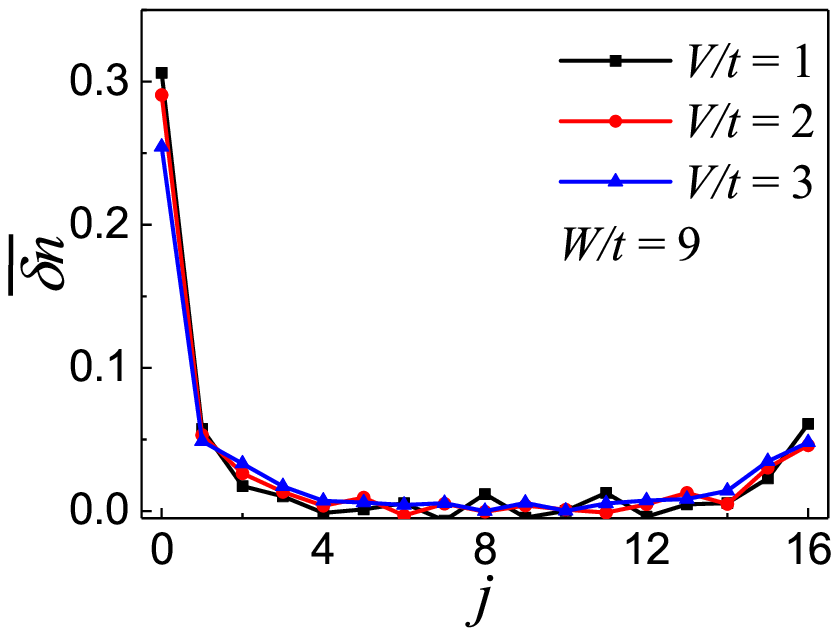}
	\includegraphics[width=0.8\linewidth]{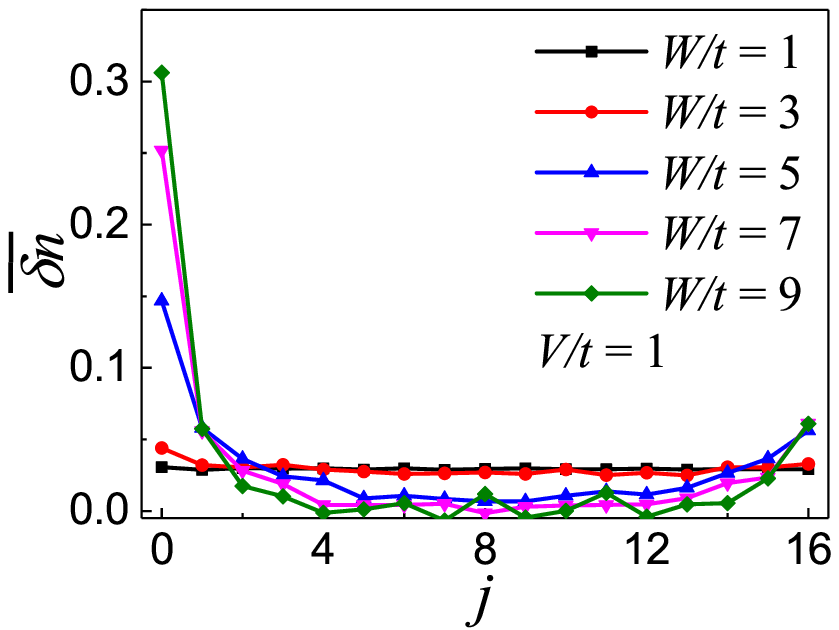}
	\caption{(Color online)
          The perturbed density profile after a local quench following protocol-(I) (see main text). 
	Here we choose the system size $L = 16$, and average over $1000$ disorder realizations in this figure. 
	(a), The density profile distribution with varying interaction strengths with disorder strength fixed at $W/t = 9$. 
	(b), The distribution for different disorder strengths with interaction fixed at $V/t =1 $.}
\label{fig:Figure1} 
\end{figure}

 To make it quantitative, here we analyze the quench dynamics following protocol-(I). We have total number of $L+1$ sites labeled from $0$ to $L$. The potential energy at site-$0$ is set to be  much higher  than other sites before quench, so is initially empty. After the quench this site gets filled in dynamics because its potential is then lowered down. Denoting the pre- and post- quench Hamiltonian as $H_0$ and $H$,  the perturbed density profile measures
$
\delta n_j = \langle \psi(\tau_1) |  {\cal O}_j  |\psi(\tau_1) \rangle 
$   
where we have 
$$
{ {\cal O}_j} = e^{iH_0 \Delta \tau} \hat{n}_j e^{-iH_0 \Delta \tau } - e^{iH \Delta \tau} \hat{n}_j e^{-iH \Delta \tau }, 
$$
with $\Delta \tau = \tau_2 - \tau_1$. 
The difference between $H$ and $H_0$ is strictly local near site-$0$. In the MBL phase, the phenomenological theory of local integrals of motion~\cite{2014_Huse_MBL_PRB,2013_Serbyn_PRL,chandran2015,Ros2015420} implies that 
the operator norm of ${\cal O}_j$ decays exponentially  
\be 
\overline{ || {\cal O} _j ||}  \propto e^{-d_j /\xi}  
\ee 
with $\xi$ a localization length, and $d_j $ the distance of the $j$-th lattice site to site-$0$, which is given by  $d_j = {\rm min}(j, L+1-j)$ in a periodic boundary system with  size $L+1$. In the thermal phase, the support of ${\cal O} _j$ would spread over the whole system through a linear light-cone dynamics. ~\cite{2012_Bloch_Light_Nature, 2017_Deng_Log_PRB, Lieb1972}
We can thus extract the localization length from the perturbed density profile $\delta n_j$ according to  
\be 
\xi = \frac{\sum_j  d_j \overline{ \delta  n_j} }{\sum_j \overline{ \delta n_j}}. 
\ee
In our numerical results shown in Fig.\ref{fig:Figure1}, we find stronger disorder strength makes the density profile more  localized and thus  $\xi$ smaller.  Increasing interaction strength makes the density profile more extended, and  $\xi$ becomes larger.



The system is completely localized at the strong disorder limit, whose eigenstates are simply product states. In the deep localized phase, the localization length is vanishing in our definition. Upon decreasing disorder, the localization length becomes larger but still remains to be a finite number in the MBL phase, i.e., with the scaling form $L^0$ as we change the system size. Further decreasing disorder, the scaling form of the localization length switch to $L/4$ immediately after the ergodic transition happens. In the one dimensional system, the scaling behavior of $\xi$ would resemble the entanglement entropy in both the localized and the ergodic phases. The difference is that $\xi$ can be measured in experiments whereas to measure the entanglement entropy is difficult. 
 
Assuming a unique diverging length scale $\delta W^{-\nu}$ at  the MBL transition leads to a natural finite-size scaling ansatz for the localization length
\be \label{eq:scaling}
\xi/L \sim  g(L^{1/\nu} \delta W)
\ee 
with $\delta W = W - W_c$. 
 
 This scaling ansatz is consistent with the fact that $\xi/L$ has a jump across the MBL-to-ergodic transition in the thermodynamic limit. 

Given the scaling form of $\xi/L$, the transition point can be extracted from the crossing of the curves for $\xi/L $ versus the disorder strength $W$ with different system sizes. 
The critical exponent $\nu$ as well as critical value $W_c$ are determined by finite-size scaling analysis~\cite{2014_Kjall_MBL_PRL,2015_Luitz_MBL_PRB}.

\begin{figure}[htp]
	\includegraphics[angle=0,width=0.8\linewidth]{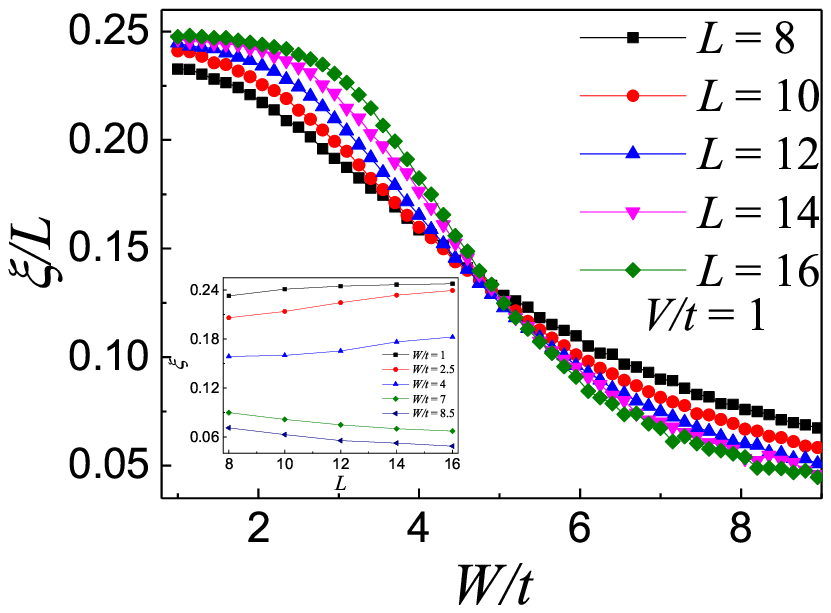}
	\quad
	\includegraphics[angle=0,width=0.8\linewidth]{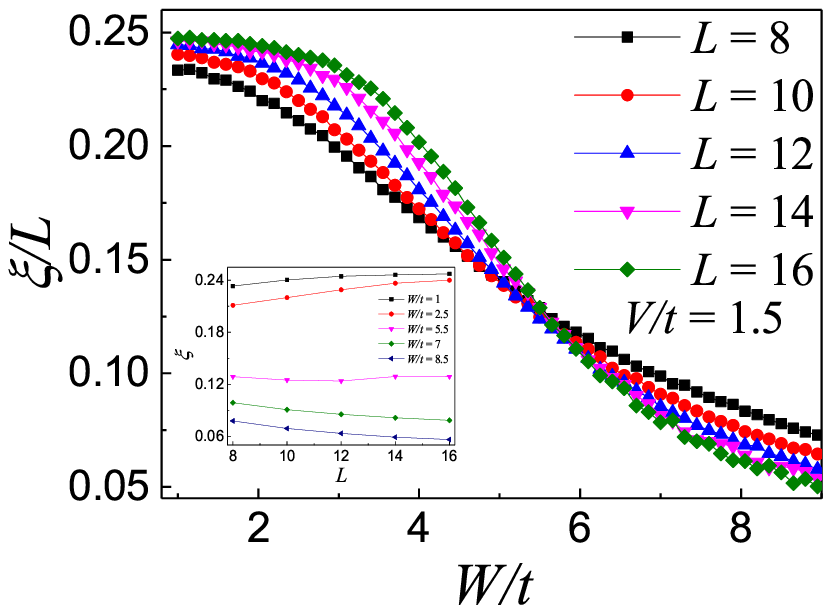}
	\quad
	\includegraphics[angle=0,width=0.8\linewidth]{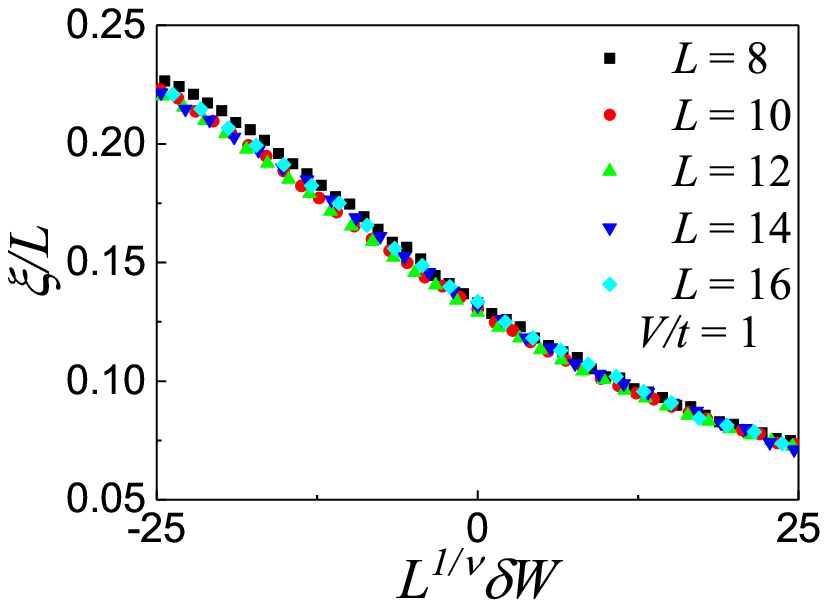}
	\caption{\label{fig:Figure2} (Color online) 
Many-body localization criticality from our proposed quench dynamics with protocol-(I) (see main text). (a) and (b) MBL length as a function of disorder strength for interaction strength $V/t = 1$ and $V/t = 1.5$ respectively. The curve intersects at $W_{c}/t = 4.90 \pm 0.03$ for $V/t = 1$ and at $W_{c}/t = 5.57 \pm 0.04$ for $V/t =1.5$. (c) Finite size scaling with scaling function $g\left( L^{1/\nu}\delta W\right )$ where $\delta W = W - W_{c}$. Critical exponent is estimated to be $\nu = 1.02 \pm 0.09$.}
\end{figure}

\section{Simulated results for post-quench dynamics} 
\label{sec:quenchresults}

The proposed dynamical experiment is simulated by performing exact diagonalization calculation. We use periodic boundary condition to minimize finite-size effects in the numerical calculation, and the system sizes are $L = 8, 10, \ldots 16$. For the experimental case, the finite-size effects are not expected to be too significant for that the number of sites in optical lattice experiments could reach the order of one hundred, much larger than that can be numerically simulated.

\subsection{Quench Protocol-(I) } 

In the quench protocol-(I), we have an empty site labeled by an index $0$, that has much higher potential energy than other lattice sites in the system. 
The pre-quench lattice sites are labeled as $1, 2, \ldots L$. In the numerical simulation of pre-quench dynamics, we restrict to the Hilbert space setting the particle number at site $0$ to be explicitly $0$. 
At $\tau_1$ the potential at site-$0$ is lowered down to the same level as other lattice sites, which instantaneously creates a hole in the system at site $0$. The system is then evolved with a long time $\tau_2$, where the created hole eventually stabilizes by interacting with other particles. 
For $L \le 12$, we perform the full matrix diagonalization and average over $10^4$ disorder realizations. For $L \ge 14$, a  Krylov space expansion is implemented for the unitary $e^{-iHt}$ in order to save cost on memory, and we average over $10^3$ disorder realizations. 
The dynamical evolution of local density $\delta n_{j} = n_{j}\left(\tau_1\right) - n_{j}\left(\tau_2\right)$  is calculated, $n_j (\tau) = \langle \psi (\tau) | \hat{n}_j |\psi(\tau)\rangle$, with $|\psi(\tau)\rangle$ the time-dependent quantum state.

Here we would like to describe a special treatment on the perturbed density profile at the quenched site, i.e. $\delta n_0$.  If we are in the thermal phase, the final state in our proposed dynamical procedure is ergodic, and the density distribution is uniform on every lattice site. 

Since at time $\tau_1$ the occupation number at site-$0$ is different from other sites, 
$\delta  n_0$ is compensated by adding $1/2$.  This compensation makes $\delta n_j$ a flat profile in the thermal phase. Meantime, the so-defined $\delta n_j$ then obeys a normalization condition  $\sum_j \delta n_j = 1/2$.

From Fig.~\ref{fig:Figure1}, averaged over different disorder realizations, $\overline {\delta n_i}$ indeed shows exponential decay in the MBL phase with strong disorder, and becomes flat in the ergodic phase at weak disorder. Note that the notation $\overline{\ldots}$ implies averaging over disorder throughout. The interaction dependence is also studied. Increasing interaction strength in the localized phase only makes the density profile of the hole a bit more delocalized, and it still shows an exponential decay, showing that the signature of many-body localization in quench dynamics is stable against interaction effects.

Fig.~\ref{fig:Figure2} (a) shows the many-body localization criticality exhibited by the hole-profile localization length in our proposed quench dynamics. 
With an interaction strength $V/t = 1$,  the extracted critical disorder strength from the localization length is found to locate at $W_{c}/t = 4.90 \pm 0.03$.  As we increase the interaction strength, the localization becomes less robust and the required disorder strength to stabilize MBL gets larger (compare Figs.~\ref{fig:Figure2} (a, b)). To find out the critical exponent $\nu$, we collapse the data to the  scaling function $g\left(L^{1/\nu}\delta W\right)$, which gives an estimate that $\nu = 1.02 \pm 0.09$. This value breaks the Harris-Chayes criterion $\nu \ge 2/d$ with $d$ the spatial dimension~\cite{1974_Harris_Effect_JPC,1986_Chayes_Finite_PRL}.  
 This violation would imply many-body localization criticality is beyond the description of conventional field theory or renormalization group study for disorder systems. 

With the numerical results, we explicitly confirm that the proposed quench dynamics can be used to study localization length and MBL criticality.

\subsection{Quench Protocol---(II)}

In the quench protocol-(II), we still let the initial state evolve for long enough time $\tau_1$. The quench is performed in a different way from protocol-(I). At $\tau_1$, we  perform a  conditional measurement on a given site, here labeled as site-$1$. Then we do a post-selection, where measurement-outcome state is discarded if there is no particle on the measured lattice site, and the state is kept otherwise. For the kept state, we remove the particle at this site to create a hole in the density profile. This process can be carried out by quantum microscope techniques in a straightforward way. For the initial state, we choose both CDW and random states---the random state case is used as a comparison to determine whether there is any artifact due to special choice of CDW states. CDW initial states have been used in cold atom experiments to study many-body localization~\cite{blochmbl}. 

\begin{figure}[htp]
	\centering
	\includegraphics[angle=0,width=0.8\linewidth]{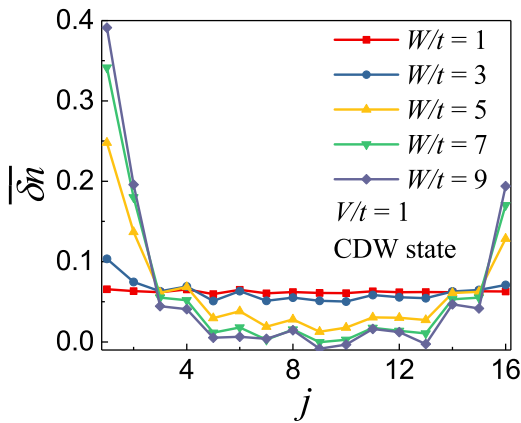}
	\caption{(Color online)
		The density profile after removing a particle on site-$1$ for different disorder strengths following quench protocol-(II) (see main text). We choose  a system size $L = 16$, and an interaction strength $V/t = 1$, and average over 1000 disorder realizations in this figure. The density profile in this plot is normalized to $1$. }
	\label{fig:Figure3} 
\end{figure}

We monitor the post-quench density profile of the hole $\delta n_j$ at a long time $\tau_{2}$. 
From Fig.~\ref{fig:Figure3}, we confirm that the hole-profile is localized (extended) in MBL (ergodic) phase. 
The local hole density is defined as $\delta n_{j} = 1/2 - n_{j}\left(\tau_{2}\right)$, and it obeys a normalization condition $\sum_j \delta n_j = 1$. 
Here, we remark that removing one particle is necessary in this quench protocol to study MBL localization length---a conditional measurement  without removing the particle could not create a well-defined hole in the density profile. 

\begin{figure}[H]
 	\centering
 	\subfigure[]{
 		\begin{minipage}{0.48\linewidth}
 			\centering
 			\includegraphics[width=1.7in]{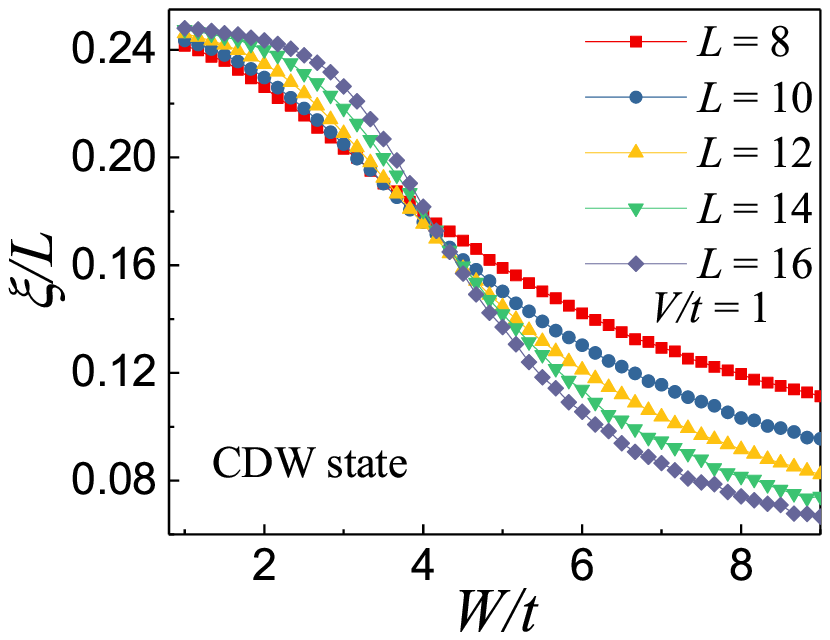}
 	\end{minipage}}
 	\subfigure[]{
 		\begin{minipage}{0.48\linewidth}
 			\centering
 			\includegraphics[width=1.7in]{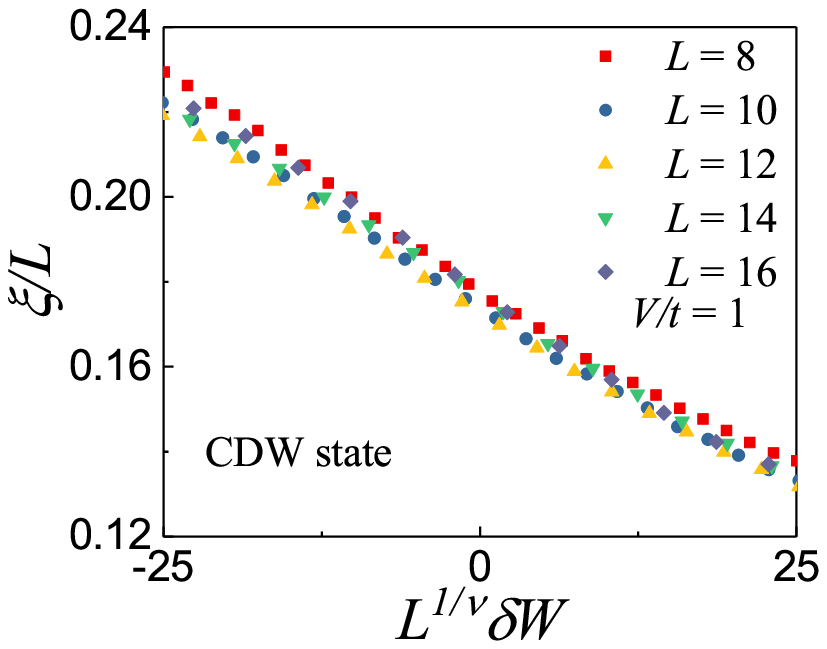}
 	\end{minipage}}
 	\quad
 	\subfigure[]{
 		\begin{minipage}{0.48\linewidth}
 			\centering
 			\includegraphics[width=1.7in]{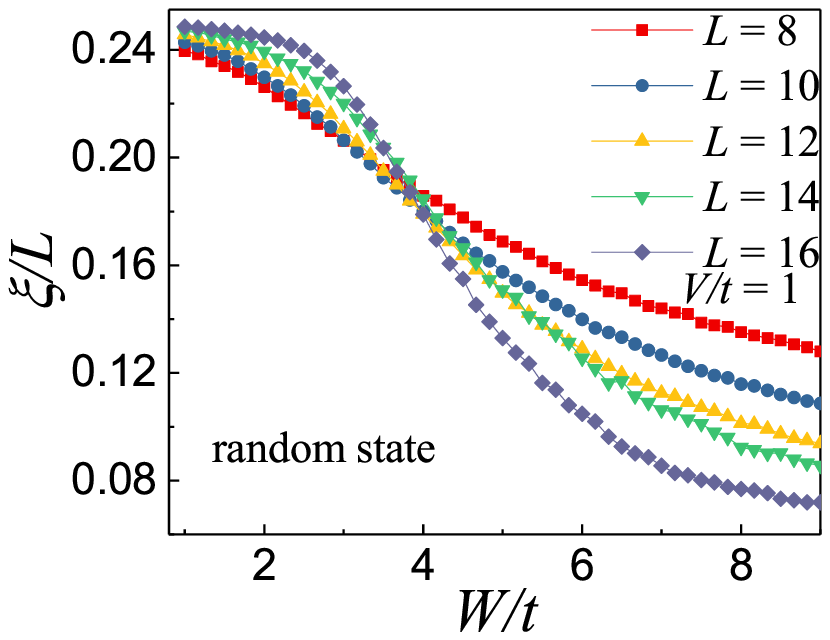}
 	\end{minipage}}
 	\subfigure[]{
 		\begin{minipage}{0.48\linewidth}
 			\centering
 			\includegraphics[width=1.7in]{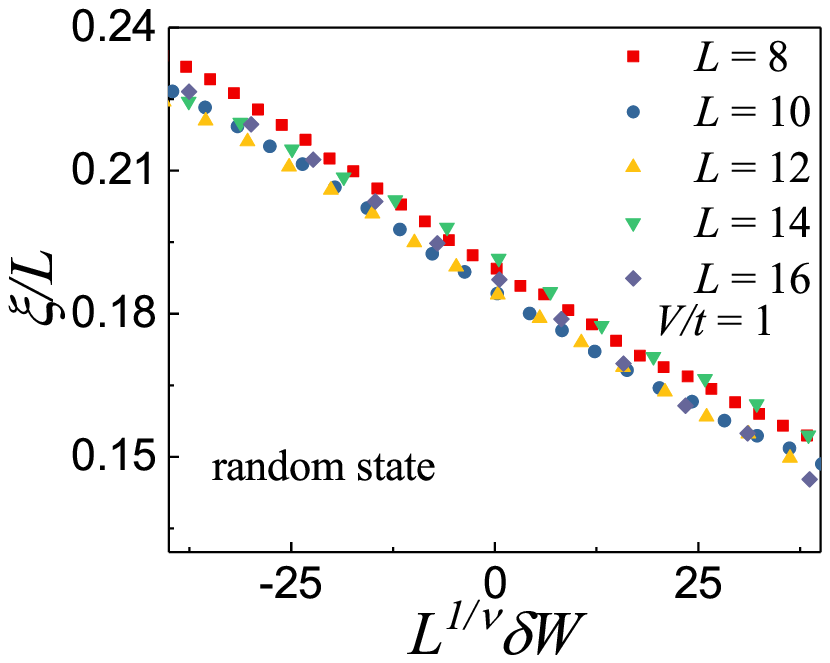}
 	\end{minipage}}
 	\subfigure[]{
 		\begin{minipage}{0.48\linewidth}
 			\centering
 			\includegraphics[width=1.7in]{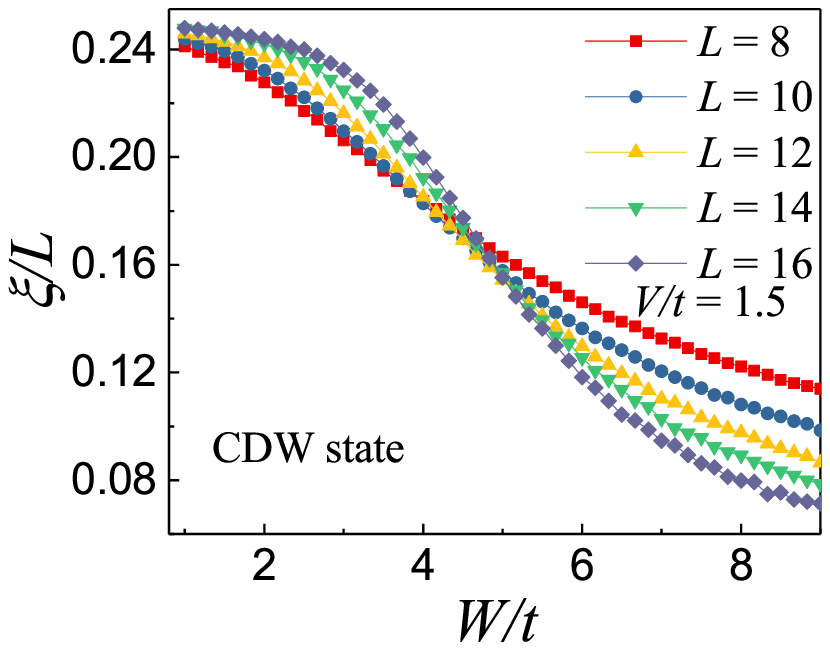}
 	\end{minipage}}
 	\subfigure[]{
 		\begin{minipage}{0.48\linewidth}
 			\centering
 			\includegraphics[width=1.7in]{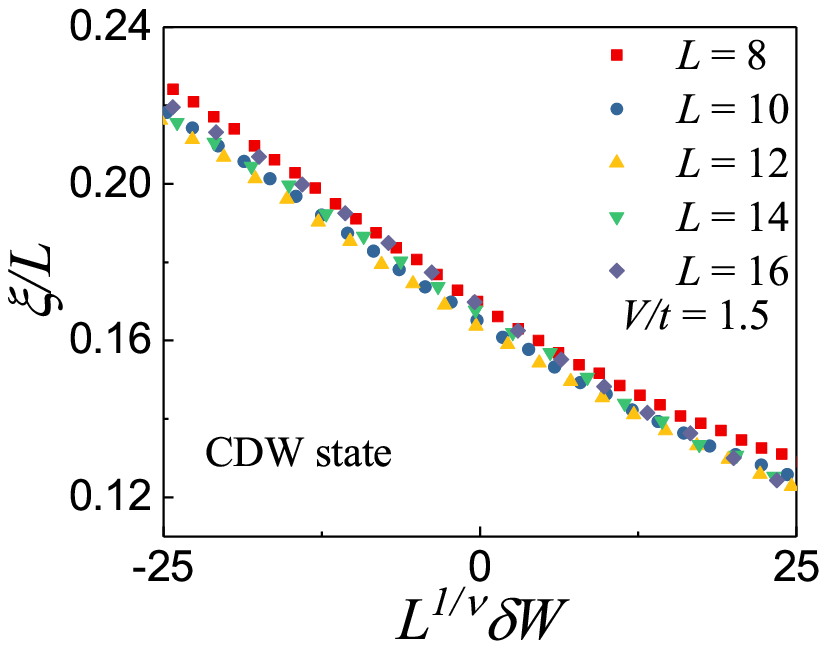}
 	\end{minipage}}
 	
 	\caption{ (Color online) 
 		Many-body localization criticality out of the quench protocol-(II) (see main text). (a, b, c, d) show the localization length versus disorder strength with the interaction fixed at $V/t = 1$. The transition point and criticality is estimated to be $W_{c}/t = 4.08 \pm 0.04$ and $\nu = 0.86 \pm 0.06$ for CDW initial states and $W_{c}/t = 3.82 \pm 0.04$ and $\nu = 0.73 \pm 0.06$ for random initial state. In (e) and (f), interaction strength is $V/t = 1.5$. The transition point and criticality is estimated to be $W_{c}/t = 4.69 \pm 0.05$ and $\nu = 0.92 \pm 0.08$. In the calculation, for $L \le 12$, the full matrix diagonalization is performed and the results are averaged over $10^4$ disorder realizations. For $L \ge 14$, a Krylov space expansion is implemented for the unitary $e^{-iHt}$, and we average over $10^3$ disorder realizations.  }
 	\label{fig:Figure4}
 \end{figure}

The localization length extracted from this protocol is shown in Fig.~\ref{fig:Figure4}.  We find its behavior is similar to the results for quench protocol-(I) (see Fig.~\ref{fig:Figure2}). 
In comparison of  CDW with random initial states,  we find no qualitative difference.  For  these two different choices of inital states, the crossing point of MBL localization length  versus disorder strength with different system sizes is consistent with each other. We find systematic data-collapse using the scaling form in Eq.~\eqref{eq:scaling} for both of them.    
This implies the MBL criticality can be studied by using CDW initial states which is relatively simpler to implement in optical lattice experiments.

Comparing the results in  Fig.~\ref{fig:Figure2} and Fig.~\ref{fig:Figure4}, it is evident that quench protocol-(II) works as well as quench protocol-(I), although the dynamics is non-unitary for (II) but unitary for (I). The similarity suggests that the details of the quench protocol are not crucial for  the study of localization length and MBL criticality, provided that the quench creates a well-defined hole in the density profile.

\begin{figure}
	\subfigure[]{
		\begin{minipage}{0.48\linewidth}
			\centering
			\includegraphics[width=1.7in]{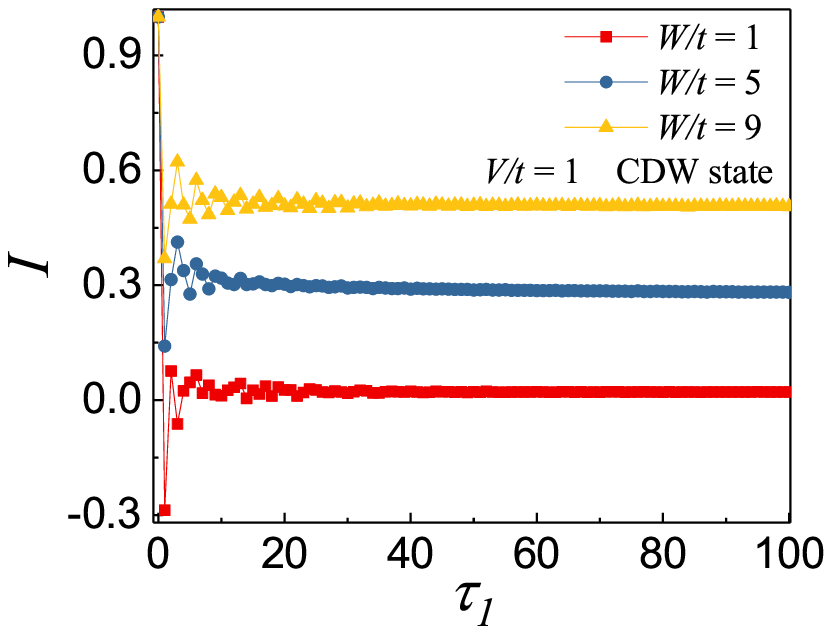}
	\end{minipage}}
	\subfigure[]{
		\begin{minipage}{0.48\linewidth}
			\centering
			\includegraphics[width=1.7in]{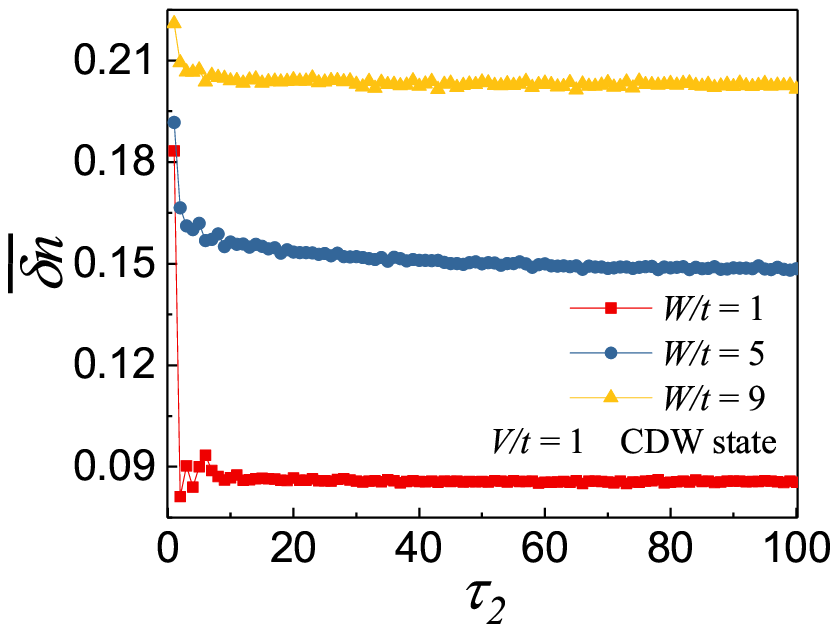}
	\end{minipage}}
	\quad
	\subfigure[]{
		\begin{minipage}{0.48\linewidth}
			\centering
			\includegraphics[width=1.7in]{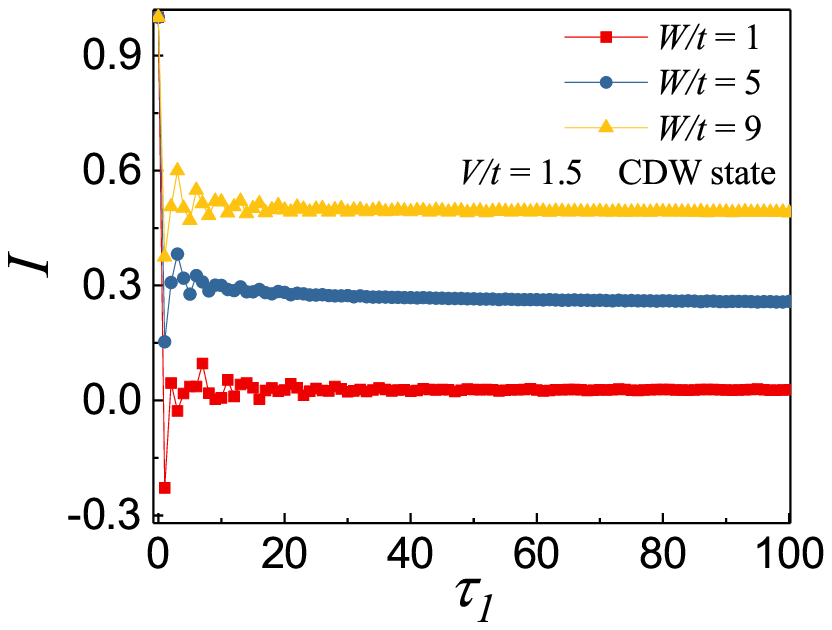}
	\end{minipage}}
	\subfigure[]{
		\begin{minipage}{0.48\linewidth}
			\centering
			\includegraphics[width=1.7in]{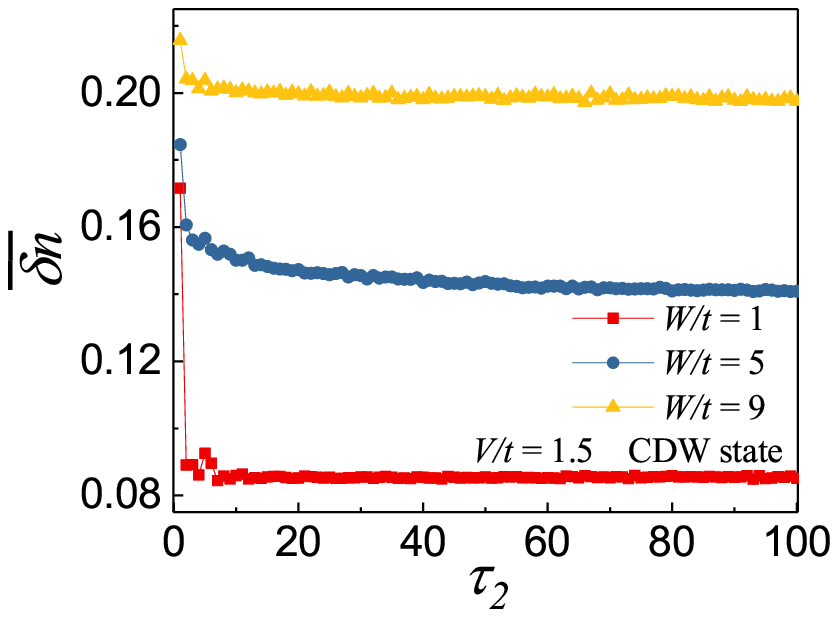}
	\end{minipage}}
	\caption{\label{fig:Figure5} 
Pre- and post-quench dynamics following the quench protocol-(II). Here we take CDW initial states, and choose $L = 12$. (a) and (c) correspond to pre-quench dynamics in number imbalance $I$ (see main text) with $V/t = 1$ and $1.5$ respectively. 
(b) and (d) show the post-quench dynamics in the density profile $\delta n_2$ for interaction strengths $V/t = 1$ and $1.5$.}
\end{figure}

In our proposal, it is crucial to know the required dynamical time scales before and after the quench, i.e., $\tau_1$ and $\tau_2$.  
For the dynamics to reveal intrinsic MBL physics, it is necessary that the system stabilizes for both before and after the quantum quench. 
In Fig.~\ref{fig:Figure5},  we show details of the dynamical evolution. 
Fig.~\ref{fig:Figure5} (a, c) show the pre-quench dynamics with $V/t = 1$ and $1.5$. We take one type of CDW state with atoms occupying all odd sites and monitor the atom number imbalance  as $I = \frac{N_{o}-N_{e}}{N_{o}+N_{e}}$, with $N_o$ and $N_e$ the total particle numbers on odd and even lattice sites.
 It can be seen that the system stabilizes after about $20$ times of tunneling time. 
Fig.~\ref{fig:Figure5} (b, d) show the post-quench dynamics in the hole density profile $\delta n_j$. 
The system is found to stabilize after about $10$ times tunneling time. 
Taking the two steps into account, the required time scale to perform the quench experiment is around $30$ times tunneling time, which is about $30-100$ milliseconds for a typical optical lattice with  tunneling time around one millisecond~\cite{2008_Bloch_Dalibard_RMP}. This is reasonably within the lifetime of cold atom experiments. 
 
%

\subsection{Interaction dependence of the critical disorder strength from local quench dynamics} 
\begin{figure} [htp]
	\includegraphics[angle=0,width=.7\linewidth]{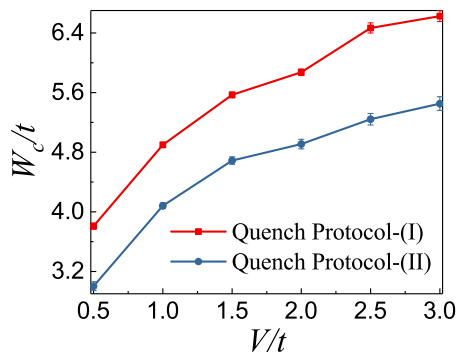}
	\caption{\label{fig:Figure7} {{Critical disorder strength as a function of interaction strength as extracted from our proposed local quench dynamics.}}}
\end{figure}

Fig.~\ref{fig:Figure7} shows a systematic study of interaction effects on MBL transition for both quench protocols.  At the interaction strength $V/t = 2$, the fermion lattice model maps onto  random field Heisenberg model which has been extensively studied in the literature. The transition point determined from our proposed dynamical experiment agrees with previous studies using other diagnostics~\cite{oganesyan2007,pal2010,moore2012,2013_Bauer_JSM,vadim,bardarson2014,2015_Singh_MBL_PRL,2015_Khemani_Pollmann_PRL,2015_Yu_Pekker_arXiv,2015_Lim_Sheng_PRB,2016_Kennes_Karrasch_PRB}. Our approach has an advantage in that all the quantities required to extract the localization transition and criticality can be probed directly in optical lattice experiments. 


It is worth noting that for large system size (see Fig.\ref{fig:Figure2}), the data crossing to determine the localization transition point has a slight rightward drift as a result of finite-size effect. To reach a conclusive answer for MBL criticality would rely on the experiments which can go to large system sizes beyond the computation capability of numerical simulations with classical resources. The advantage of our proposed strategy to probe MBL criticality is that the required ingredients are all presently accessible with cold atoms in optical lattices. 

\section{Benchmarking the dynamical protocol with Anderson localization}

\begin{figure} [H]
	\centering
	\subfigure[]{
		\begin{minipage}{0.48\linewidth}
			\centering
			\includegraphics[width=1.7in]{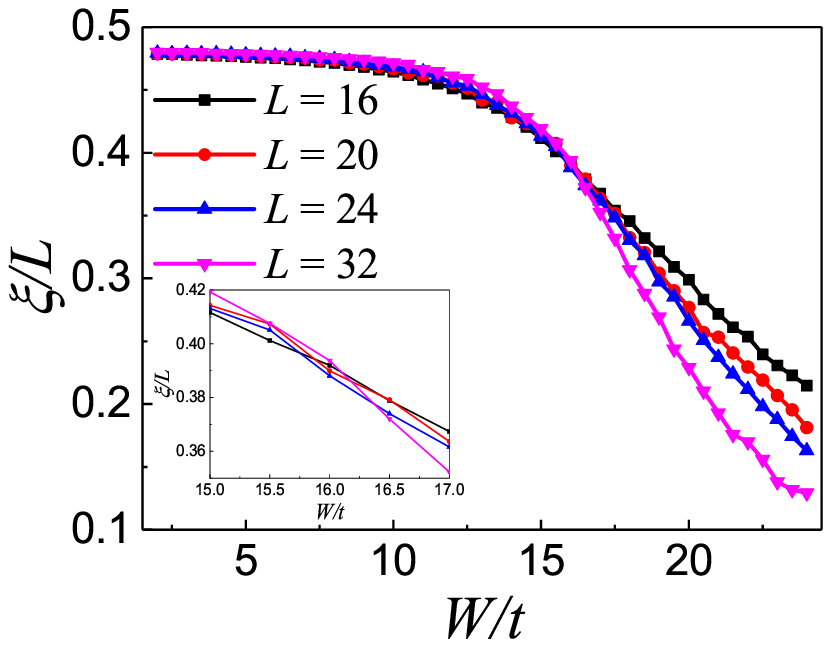}
	\end{minipage}}
	\subfigure[]{
		\begin{minipage}{0.48\linewidth}
			\centering
			\includegraphics[width=1.7in]{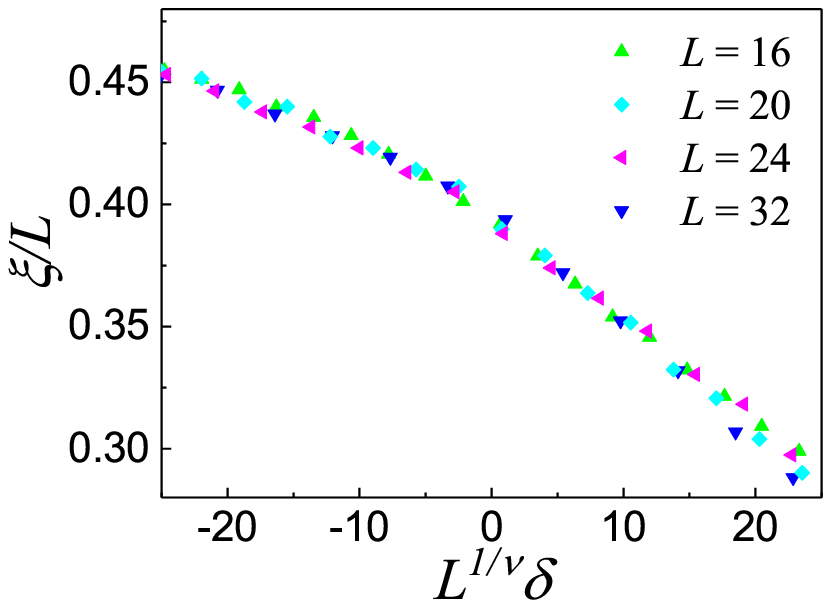}
	\end{minipage}}
	\caption{\label{fig:Figure8} (Color online) (a) Localization length as a function of disorder strength for the three dimensional Anderson model to benchmark our method. The curves for different system sizes intersects at $W_c/t = 15.88 \pm 0.14$ . (b) Data collapse to a  scaling function $g_{\rm AL} \left[L^{1/\nu}\delta\right]$ where $\delta = W - W_{c}$, which leads to  $\nu = 1.6 \pm 0.2$.}
\end{figure}

To benchmark the method of probing the many-body localization length in our study,  we carry out a simulation for the well-understood three-dimensional Anderson localization whose Hamiltonian reads, 
\begin{equation} \label{Hamiltonian2}
\hat{H} = \sum_{<jj'>}-t\left(\hat{c}_{j}^{\dagger}\hat{c}_{j'}+h.c.\right) + \sum_{j}h_{j}\hat{n}_{j}. 
\end{equation} 
For this non-interacting model, we extract the localization length from the long-time-evolved density profile of a single particle initialized at one lattice site. This single-particle  density profile indeed takes an exponential decay form. For $L = 16$ ($20$, $24$ and $32$), the density profile is averaged over $1000$ ($100$) realizations and in the vicinity of expected transition interval, we average over 1000 realizations for all sizes. 
By calculating the localization length from the density profile, we find $W_c/t = 15.88 \pm 0.14$ (see Fig.~\ref{fig:Figure8}) and the critical exponent $\nu = 1.6 \pm 0.2$, which are consistent with well-known results for the Anderson model~\cite{1999_Slevin_Corrections_PRL}.  This justifies the validity of our proposed dynamical protocols to extract localization criticality. 



\section{Conclusion} 

In summary, we propose to use local quench dynamics to probe the MBL-to-ergodic criticality. Its validity is confirmed by benchmarking towards  the well-known 3D Anderson localization. In our proposed schemes, the localization length could be extracted from the exponential decay of a perturbed density profile after a local quantum quench. This proposal is expected to guide future experiments in probing MBL criticality with ultracold atoms in disordered optical lattices. Moreover, we provide a finite-size scaling form of the localization length across the transition, which is directly applicable in analyzing the quench-dynamics data out of the proposed experiments. It is worth future study to sort out the finite-size effects in the quench dynamics of many-body localization, in particular about violation of Harris-Chayes bound, for example by developing dynamical renormalization group techniques.

\section{Acknowledgments}
{We thank I. Bloch, Y. Takahashi, D.A. Huse, E. Altman, G. Refael, X. Chen, J. Alicea, L. Jiang and S. M. Girvin for helpful discussions. 
This work is supported by National Program on Key Basic Research Project of China under Grant No. 2017YFA0304204, National Natural Science Foundation of China under Grants No. 117740067,  and the Thousand-Youth-Talent Program of China. 
}

\bibliography{references}
\bibliographystyle{apsrev4-1}

\end{document}